\documentclass[%
 reprint,
 superscriptaddress,
 amsmath,amssymb,
pra,
longbibliography
]{revtex4-1}

\usepackage{graphicx}
\usepackage{dcolumn}
\usepackage{bm}
\usepackage{braket} 
\usepackage[caption=false]{subfig} 
\usepackage{changepage}
\usepackage{float}
\usepackage{amsmath}
\usepackage{xcolor}
\usepackage{soul} 
\usepackage{tikz}
 
\usepackage{textgreek}

\begin{document}


\title{Lorentz invariant materials and metamaterials}

\author{Jon Lasa-Alonso}
\altaffiliation[jlasa@mondragon.edu]{}
\affiliation{Basic Sciences Department, Mondragon Unibertsitatea, Olagorta kalea 26, 48014 Bilbao, Spain}

\author{Gabriel Molina-Terriza}
\affiliation{Donostia International Physics Center, Paseo Manuel de Lardizabal 4, 20018 Donostia-San Sebasti\'an, Spain}
\affiliation{Centro de F\'isica de Materiales (CSIC-UPV/EHU), Paseo Manuel de Lardizabal 5, 20018 Donostia-San Sebasti\'an, Spain}
\affiliation{IKERBASQUE, Basque Foundation for Science, Mar\'ia D\'iaz de Haro 3, 48013 Bilbao, Spain}

\date{\today}

\begin{abstract}
We show that electromagnetic constitutive relations of materials, defined by a scalar permittivity, $\varepsilon$, and permeability, $\mu$, are the same for all inertial observers provided that $\varepsilon\mu = 1$ is satisfied. To highlight the practical power of our general result, we discuss schemes in which this property plays a significant role, such as Fresnel-Fizeau optical drag or modified Snell's law for moving dielectrics. We also proof that such a particular behavior can be extended to systems like uniaxial anisotropic media or negative refractive index metamaterials.

\end{abstract}

\maketitle


In macroscopic electrodynamics, many problems consider the presence of materials whose properties are classified in terms of symmetries.

For instance, we say that a macroscopic sample is homogeneous whenever its composition does not change from one point of space to another. We may also say that a material is isotropic, provided that its electromagnetic response does not change under rotations. In addition, if the properties of a sample do not vary over time, we most commonly say that the material is static. In contrast, samples can also be identified for not having the aforementioned properties, i.e., for being inhomogeneous, anisotropic, or time-varying. All of these features are translated into specific constraints for the tensors defining the electromagnetic constitutive relations. In the local and linear response regime, non-chiral materials are defined by two tensor functions, i.e., the relative permittivity tensor, $\overline{\varepsilon}(\mathbf{r}, t)$, and the relative permeability tensor, $\overline{\mu}(\mathbf{r}, t)$, where $\mathbf{r}$ is the position vector and $t$ the time. If the material is identified as homogeneous and static, it means that we can drop the dependence on $\mathbf{r}$ and $t$, respectively. On the other hand, if we are dealing with isotropic materials, we can express the constitutive relations in terms of scalar permittivity and permeability functions. A material that is simultaneously homogeneous, isotropic and static can be represented by two independent magnitudes, i.e., the scalar permittivity, $\varepsilon$, and the scalar permeability, $\mu$.

Materials can be identified as being invariant under different symmetry groups. In particular, an environment described by constant $\varepsilon$ and $\mu$ parameters, is invariant under the group containing continuous spatial translations, continuous spatial rotations and continuous time translations. However, it is a well-known result of Classical Electrodynamics and Special Relativity that such a system is not invariant under Lorentz transformations \cite{Mink_original}. In physical terms, this implies that the electromagnetic response of materials is different when observed from distinct inertial reference frames. Or, equivalently, we say that the constitutive relations are modified when samples move with certain speed with respect to the laboratory frame. In this Letter, we show that this is overcome by materials for which the relation of the scalar permittivity and permeability is $\varepsilon = 1/\mu$, independently of the value of $\mu$. Notice that these media include, of course, the vacuum, but are not limited to it.  Unless stated otherwise, we restrict ourselves to homogeneous, isotropic, static materials, to simplify derivations and focus on the physical phenomena. We first prove that constitutive relations of dielectric materials fulfilling this constraint are left unchanged when they move at an arbitrary speed. Then, we show that this particular behavior can also be met in anisotropic media and negative index metamaterials. Finally, we discuss specific problems in which Lorentz invariant materials play a significant role, such as Fresnel-Fizeau optical drag or modified Snell's law for moving dielectrics.

\begin{figure}[t]
    \centering
    \includegraphics[width = 0.5\textwidth]{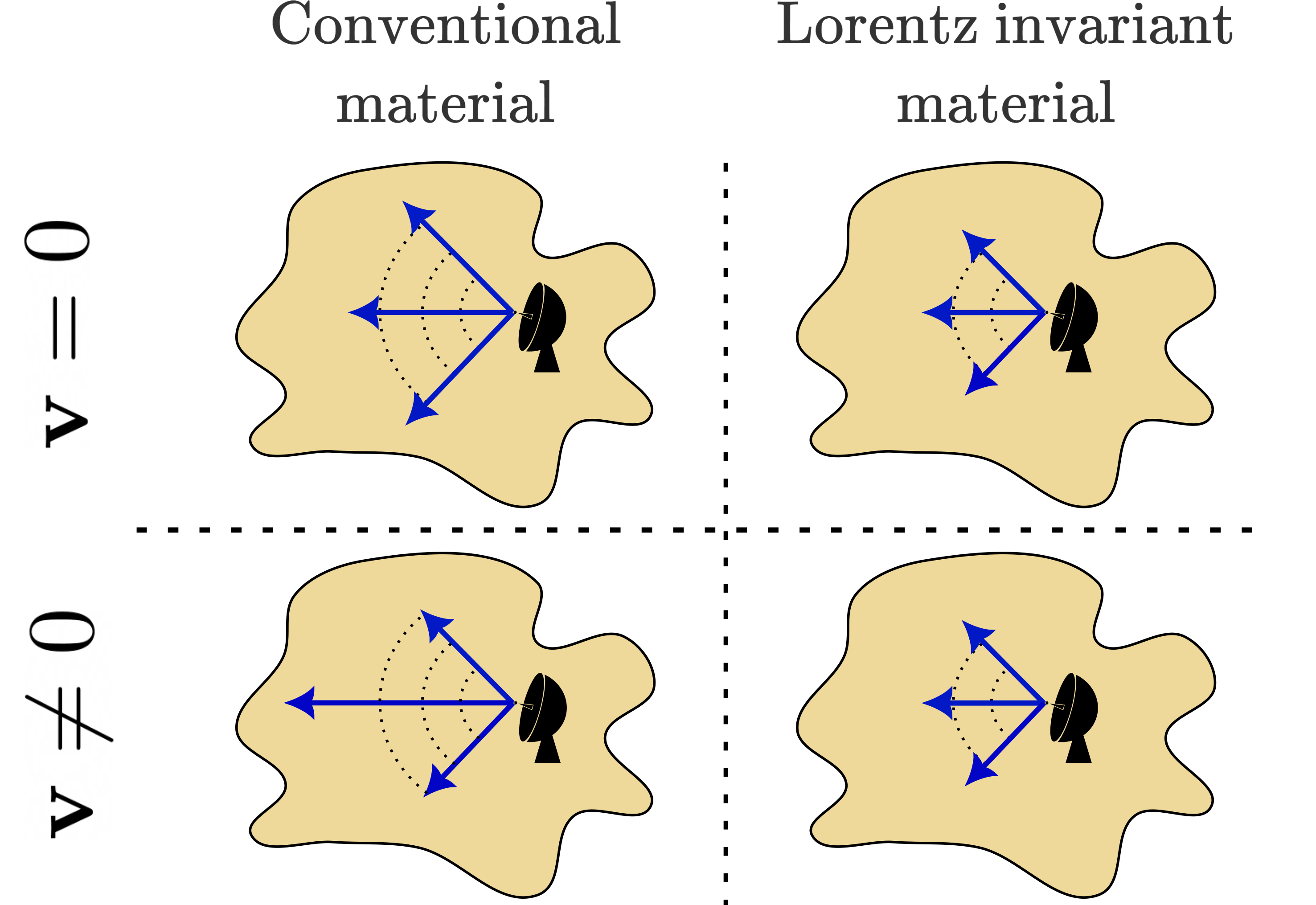}
    \caption{Electromagnetic waves propagating through media which are moving ($\mathbf{v} \neq 0$) or not moving ($\mathbf{v} = 0$) with respect to an emitter placed inside (laboratory frame). Electromagnetic response of conventional materials is modified when they move at a certain speed $\mathbf{v}\neq 0$. This is depicted on the left column, where an emitter inside the medium would observe a difference in the propagation of electromagnetic waves, when the medium is moving compared to when it is at rest. For a Lorentz invariant material, electromagnetic properties are left unchanged. As a result, an emitter inside would not observe any difference in the propagation of electromagnetic waves.}
    \label{Fig_intro}
\end{figure}

Let us first briefly discuss how a Lorentz transformation is performed over a generic dielectric system. From a passive point of view, performing such a transformation implies studying the sample from an inertial reference frame moving at a constant speed $-\mathbf{v}$, with respect to the co-moving frame. Here, the modulus of the velocity has to fulfill that $|\mathbf{v}| < 1$ if we consider natural units, i.e., that the speed of light is $c=1$. Note that this situation is equivalent to keeping the observer at rest and considering the sample moving at speed $\mathbf{v}$. The latter way of addressing a Lorentz transformation is usually denoted as the active perspective, and it is the one we employ here. Therefore, considering all possible Lorentz transformations over a dielectric system is equivalent to studying the behavior of the probe when it is moving at an arbitrary constant speed $\mathbf{v}$ with respect to the laboratory frame. In this line, a Lorentz invariant dielectric material is one whose electromagnetic properties are independent of speed. A sample whose electromagnetic response does not change with speed is perceived equally by all inertial reference frames and, thus, it can be regarded as being invariant under Lorentz transformations (Fig. \ref{Fig_intro}).

To prove that a dielectric system is Lorentz invariant, we must check how the set of Maxwell's equations 
\begin{align}
    \nonumber
	&\nabla \times \mathcal{E}(\mathbf{r}, t) = -\partial_t \mathcal{B}(\mathbf{r}, t),~~~~~
	\nabla \cdot \mathcal{D}(\mathbf{r}, t) = 0\\
    \label{Maxwell}
	&\nabla \times \mathcal{H}(\mathbf{r}, t) = \partial_t\mathcal{D}(\mathbf{r}, t),~~~~~~
	\nabla \cdot \mathcal{B}(\mathbf{r}, t) = 0
\end{align}
together with the constitutive relations
\begin{equation}
    \label{CRel}
    \mathcal{D}(\mathbf{r}, t) = \varepsilon \mathcal{E}(\mathbf{r}, t),~~~~\mathcal{B}(\mathbf{r}, t) = \mu\mathcal{H}(\mathbf{r}, t),
\end{equation}
are modified when considering a sample moving at constant speed $\mathbf{v}$. This specific problem was addressed by Minkowski in 1908 \cite{Mink_original, Einstein_Laub, Minkowski}, and he found that whereas the form of Maxwell's equations as given in Eq. \eqref{Maxwell} does not change for moving bodies, the constitutive relations expressed in Eq. \eqref{CRel} must be modified. The form that constitutive relations acquire for a body moving at constant speed $\mathbf{v}$ is \cite{Landau}:
\begin{align}
    \label{CRel_v1}
    \mathcal{D} + \mathbf{v}\times\mathcal{H} &= \varepsilon \left(\mathcal{E} + \mathbf{v}\times\mathcal{B} \right)\\
    \label{CRel_v2}
    \mathcal{B} - \mathbf{v}\times\mathcal{E} &= \mu \left(\mathcal{H} - \mathbf{v}\times\mathcal{D} \right),
\end{align}
where ``$\times$" represents the vectorial product and, for clarity, $(\mathbf{r}, t)$ dependence of the fields has been omitted. Equations \eqref{CRel_v1} and \eqref{CRel_v2} indicate that a probe which is isotropic in the co-moving frame appears to be bianisotropic for an inertial reference frame moving at a constant speed $-\mathbf{v}$ \cite{Sommerfeld}.

In what follows we show that there exists a subset of dielectric samples for which the constitutive relations at rest, as specified by Eq. \eqref{CRel}, are left unchanged when they move with an arbitrary constant speed $\mathbf{v}$. To do so, we switch to an alternative representation of the electromagnetic field, in terms of the Riemann-Silberstein (RS) vector \cite{RoleRS_Bialynicki-Birula} (for an equivalent proof based on the conventional electric and magnetic fields see Supplementary Material). For those readers that may not be familiarized, we note that the RS vector constitutes the electromagnetic version of the so-called Beltrami fields, commonly employed in  fluid mechanics \cite{Lakhtakia_Beltrami}.

The RS vector may be defined in a variety of ways, but here we take it to be constructed as:
\begin{align}
    \label{F}
    \mathcal{F}(\mathbf{r}, t) = \frac{1}{\sqrt{2}}\left[ \frac{\mathcal{D}(\mathbf{r}, t)}{\sqrt{\varepsilon}} + i \frac{\mathcal{B}(\mathbf{r}, t)}{\sqrt{\mu}} \right],
\end{align}
and also a version defined in terms of the fields $\mathcal{E}$ and $\mathcal{H}$:
\begin{align}
    \label{G}
    \mathcal{G}(\mathbf{r}, t) = \frac{1}{\sqrt{2}}\left[ \frac{\mathcal{E}(\mathbf{r}, t)}{\sqrt{\mu}} + i \frac{\mathcal{H}(\mathbf{r}, t)}{\sqrt{\varepsilon}} \right],
\end{align}
with $i=\sqrt{-1}$. At this point, it is important to note that the linear combinations in Eqs. \eqref{F} and \eqref{G} consider real valued $\mathcal{D}$, $\mathcal{B}$, $\mathcal{E}$ and $\mathcal{H}$ fields. Thus, $\varepsilon$ and $\mu$ must also be real numbers and, in addition, we first consider that they are both positive, i.e., $\varepsilon > 0$ and $\mu > 0$. In terms of the RS vector, the constitutive relations for a dielectric sample at rest $(\mathbf{v} = 0)$ can be expressed simply as: $\mathcal{F} = n\mathcal{G}$, where we have defined the refractive index of the sample to be $n = \sqrt{\varepsilon\mu}$ and both relations specified in Eq. \eqref{CRel} can be retrieved by taking the real and imaginary parts. On the other hand, we can also employ the RS vector to represent the constitutive relations for a macroscopic body moving at constant speed $\mathbf{v}$:
\begin{equation}
    \label{CRel_vF}
    \mathcal{F} - i\mathbf{v}\times\mathcal{G} = n\mathcal{G} -i n \mathbf{v}\times\mathcal{F}.
\end{equation}
In this case, Eqs. \eqref{CRel_v1} and \eqref{CRel_v2} are retrieved by considering the real and imaginary parts of Eq. \eqref{CRel_vF}, respectively.

Let us have a closer look to the form of the constitutive relations for a moving body. Indeed, by rearranging the terms in Eq. \eqref{CRel_vF}, we can express them as:
\begin{equation}
    \label{CRel_vFmat}
    \Big[ \overline{\text{I}} + in\overline{\text{V}} \Big]\mathcal{F} = \Big[ \overline{\text{I}} + \frac{i}{n}\overline{\text{V}} \Big]n\mathcal{G},
\end{equation}
where $\overline{\text{I}}$ represents the three dimensional identity operator and we have defined $\overline{\text{V}} = \mathbf{v}\times$. Note that operators acting on both sides of Eq. \eqref{CRel_vFmat} can be expressed as matrices whose determinants cannot vanish simultaneously. This implies that for every dielectric system and regardless of the speed they may carry, we can always compute the inverse of at least one of the two operators appearing in Eq. \eqref{CRel_vFmat}. As a result, and with any loss of generality, we can express the constitutive relations of a dielectric probe moving at a constant speed $\mathbf{v}$ as:
\begin{equation}
\label{CRel_vF_final}
\begin{cases}
\begin{array}{ll} \mathcal{F} = (\overline{\text{O}}^{\scriptscriptstyle{-1}}\overline{\text{P}})n\mathcal{G} \quad\quad \text{if} ~~ n^2|\mathbf{v}|^2 \neq 1 \\ (\overline{\text{P}}^{\scriptscriptstyle{-1}}\overline{\text{O}})\mathcal{F} = n\mathcal{G} \quad\quad \text{else}, \end{array}
\end{cases}
\end{equation}
where we have defined the operators $\overline{\text{O}} = \overline{\text{I}} + in\overline{\text{V}}$ and $\overline{\text{P}} = \overline{\text{I}} + in^{\scriptscriptstyle{-1}}\overline{\text{V}}$. Note that this form of the constitutive relations, as given in Eq. \eqref{CRel_vF_final}, is equivalent to others previously reported in the literature \cite{Pauli}.

Equation \eqref{CRel_vF_final} indicates that there is exactly one possibility for the constitutive relations of a moving medium to be the same as when the medium is at rest, i.e., by considering $\overline{\text{O}} = \overline{\text{P}}$. Under this condition, $\overline{\text{O}}^{\scriptscriptstyle{-1}}\overline{\text{P}} = \overline{\text{P}}^{\scriptscriptstyle{-1}}\overline{\text{O}} = \overline{\text{I}}$ is fulfilled and, thus, $\mathcal{F} = n\mathcal{G}$ is recovered in both cases of Eq. \eqref{CRel_vF_final}. Crucially, apart from the trivial solution ($\mathbf{v} = 0$), there is only one other way of fulfilling $\overline{\text{O}} = \overline{\text{P}}$ and it is by considering a dielectric system for which $n = n^{\scriptscriptstyle{-1}}$. If we express such a constraint in terms of the original permittivity and permeabilities, we obtain that this type of macroscopic media fulfills the constraint $\varepsilon\mu = 1$. In other words, there exists a well-defined subset of dielectric materials for which the constitutive relations are left unchanged independently of the speed they carry. This constitutes the central result of our work. To the best of our knowledge, it is the first time that an exact proof of this statement is given for an arbitrary velocity of the dielectric samples.

Our result has far reaching consequences. First, we would like to note that, as mentioned earlier, vacuum is evidently included within the subset of Lorentz invariant materials. While in the case of vacuum $\varepsilon = 1$ and $\mu = 1$, note that other dielectric macroscopic systems for which $\varepsilon \neq 1$ and $\mu \neq 1$ may also fulfill the necessary constraint, i.e., that $\varepsilon = 1/\mu$. This implies that there is an infinite set of dielectric materials for which the symmetries of Special Relativity are completely retrieved. For all this class of Lorentz invariant materials the speed of electromagnetic waves propagating in such media is the same than in vacuum. In addition, we note that, while this kind of materials may be difficult to find in natural conditions, one could design them using metamaterials, and in this case we can even explore Lorentz invariant macroscopic systems with a negative refractive index. For this type of samples both the permittivity and permeability are negative, i.e., $\varepsilon < 0$ and $\mu < 0$. As a result, one must take the negative square root when considering the refractive index, i.e., $n = -\sqrt{|\varepsilon||\mu|}$ \cite{Veselago, Pendry}. However, note that the product of permittivity and permeability for a negative index material may still fulfill the constraint $\varepsilon\mu = 1$, as long as $\varepsilon = 1/\mu$, exactly as for conventional dielectric materials. 

While our proof is written for positive indices of refraction, the extension to the case of negative refractive index metamaterials is straightforward. We should just slightly modify the definition of the RS vector to: $\mathcal{F} = 2^{\scriptscriptstyle{-1/2}} \{|\varepsilon|^{\scriptscriptstyle{-1/2}}\mathcal{D} + i|\mu|^{\scriptscriptstyle{-1/2}}\mathcal{B}\}$ and, accordingly, $\mathcal{G} = 2^{\scriptscriptstyle{-1/2}} \{|\mu|^{\scriptscriptstyle{-1/2}}\mathcal{E} + i|\varepsilon|^{\scriptscriptstyle{-1/2}}\mathcal{H}\}$. This redefinitions lead to a different expression of the constitutive relations for a sample at rest, i.e., $\mathcal{F} = -\eta\mathcal{G}$, where $\eta = \sqrt{|\varepsilon||\mu|}$ and relations in Eq. \eqref{CRel} are retrieved by considering the real and imaginary parts. As a result, the expression for the constitutive relations for a sample with a negative refractive index moving at constant speed $\mathbf{v}$ is:
\begin{equation}
    \label{CRel_NRI}
    \Big[ \overline{\text{I}} - i\eta\overline{\text{V}} \Big]\mathcal{F} = -\Big[ \overline{\text{I}} - \frac{i}{\eta}\overline{\text{V}} \Big]\eta\mathcal{G}.
\end{equation}
Note that Eqs. \eqref{CRel_v1} and \eqref{CRel_v2} are also retrieved if we consider the real and imaginary parts of Eq. \eqref{CRel_NRI}, respectively. Exactly as before, there are two ways of recovering the constitutive relations at rest ($\mathcal{F}=-\eta\mathcal{G}$): either by imposing the trivial condition ($\mathbf{v} = 0$) or by restricting to negative refractive index materials for which $\eta = \eta^{\scriptscriptstyle{-1}}$. Finally, if we express this last constraint in terms of the permittivity and permeability of the material, we get that $|\varepsilon||\mu| = \varepsilon\mu = 1$ has to be fulfilled. In the same spirit, one can similarly extend the proof to anisotropic media, which cannot be described with scalar permeabilities and permittivities (see Supplementary Material).

Although it has been previously often assumed that Lorentz invariance holds only for vacuum  \cite{Science_only_vacuum}, our results prove that there is a range of dielectric materials, metamaterials and anisotropic media which can fulfill the necessary constraints. In our view, this opens up a new paradigm in which symmetries of Special Relativity can be considered for a variety of different underlying environments. In the following, we discuss schemes in which Lorentz invariant materials play a significant role. We do this to put forward the practical power that our general result has in the discussion and analysis of specific physical problems. If symmetries of Special Relativity are recovered for materials fulfilling $\varepsilon\mu = 1$, this fact needs to be present in different electromagnetic problems and magnitudes measurable for moving electromagnetic systems.

\begin{figure}[t]
    \centering
    \includegraphics[width = 0.4\textwidth]{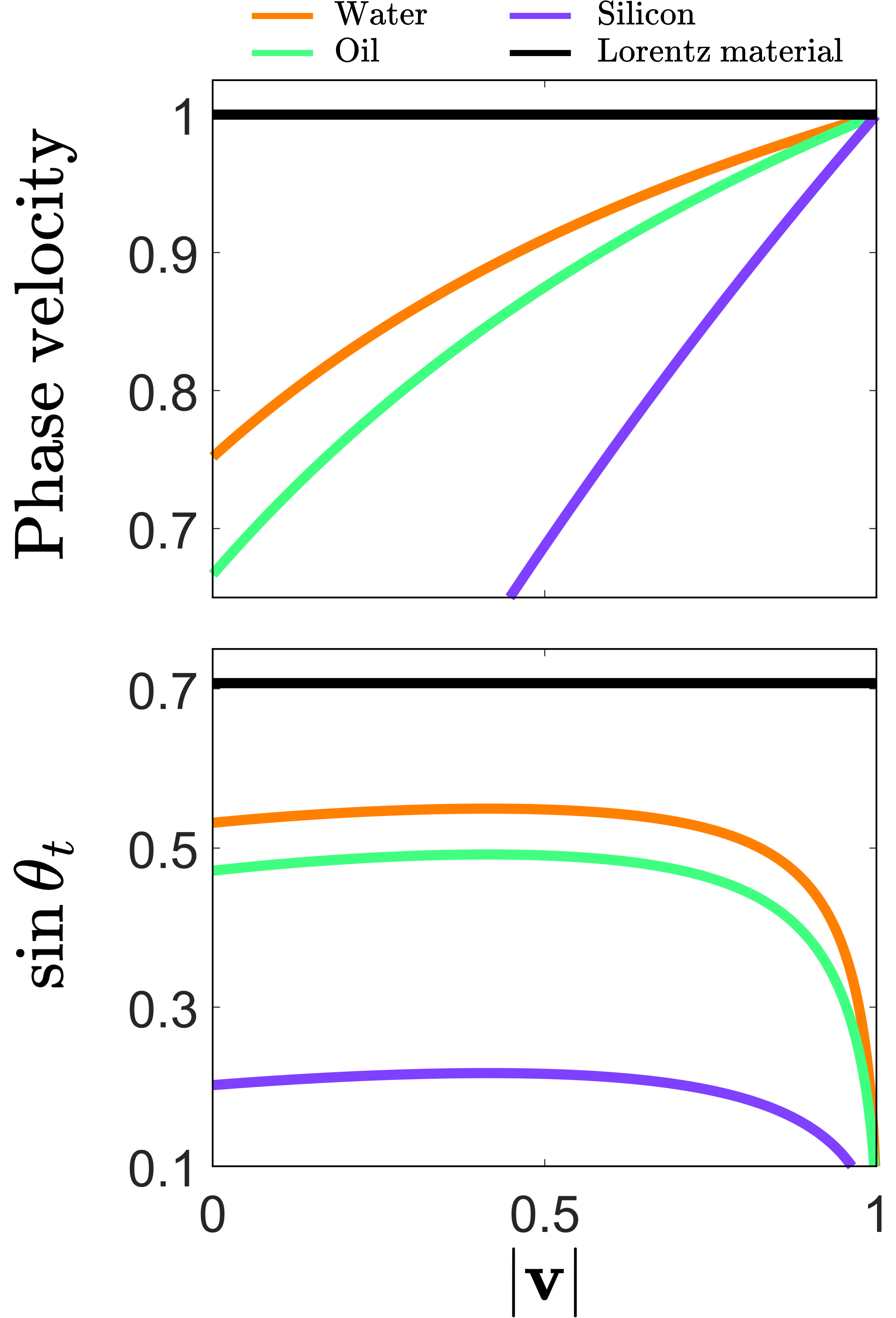}
    \caption{Upper panel: exact expression of the phase velocity for water ($\varepsilon = 1.33^2$, $\mu =1$), oil ($\varepsilon = 1.5^2$, $\mu = 1$), silicon ($\varepsilon = 3.5^2$, $\mu = 1$) and a Lorentz invariant material ($\varepsilon = 2$, $\mu = 0.5$), as a function of the medium speed modulus, $|\mathbf{v}|$ ($\psi = 0$). Lower panel: angle of refraction for the same set of materials ($\theta_i = \pi/4$, $\phi_i = \pi/2$).}
    \label{Fig1}
\end{figure}

For the rest of the manuscript, we stick to isotropic dielectric samples. The symmetries that are left in moving isotropic dielectric environments, i.e., continuous time and space translations, suggest that monochromatic plane-waves should be possible solutions of these systems. Thus, we may look for solutions of the form $\mathcal{E}(\mathbf{r}, t) = \text{Re}\left\{\mathbf{E}_0 e^{i(\mathbf{k}\cdot\mathbf{r} - \omega t)}\right\}$ and similarly for $\mathcal{H}$, $\mathcal{B}$ and $\mathcal{D}$ fields. In the expression above $\mathbf{E}_0$ represents the complex amplitude of the electric field, $\mathbf{k}$ is the wavevector and $\omega$ is the angular frequency. This constitutes the starting point to analyze the properties of monochromatic plane-waves. For instance, one can retrieve the dispersion relation of electromagnetic waves propagating through moving dielectric media. Such a derivation yields \cite{Chen}:
\begin{equation}
    \frac{|\mathbf{k}|}{\omega} = \frac{-|\mathbf{v}|\xi\cos \psi + \sqrt{1 + \xi(1 - |\mathbf{v}|^2\cos^2\psi)}}{1 - \xi |\mathbf{v}|^2\cos^2\psi},
\end{equation}
where $\xi = (\varepsilon\mu -1)\gamma^2$ with $\gamma = (1 - |\mathbf{v}|^2)^{-1/2}$, and $\psi$ is the angle between $\mathbf{k}$ and $\mathbf{v}$ vectors. It can be checked that Lorentz invariant materials leave the dispersion relation unchanged regardless of the speed of the medium, and it can be proved that they are the only class of materials for which this happens (see Supplementary Material). In particular, the well-known Fresnel-Fizeau optical drag of the phase velocity ($v_p$) \cite{Fresnel_drag, Fizeau_drag}, which is calculated for small medium velocities, i.e.,
\begin{equation}
    v_p \sim \frac{1}{\sqrt{\varepsilon\mu}} + \left( 1 - \frac{1}{\varepsilon\mu} \right)|\mathbf{v}|\cos\psi,
\end{equation}
would also disappear for Lorentz invariant materials. For all ranges of the speed of the medium, the only environments for which $v_p$ is not altered are, precisely, Lorentz invariant materials (Fig. \ref{Fig1}, upper panel). 

We have also examined the effect over the reflection and refraction of plane-waves encountering a semi-infinite plane moving at a constant speed, $\mathbf{v}$. This is an example in which one can observe that while Lorentz invariant media have properties similar to those of vacuum, they are different than vacuum. Let us take the situation where the speed of the medium is parallel to the planar interface \cite{Pyati}. In general, while the angle of refraction changes with the angle of incidence and the speed of the medium (Fig. \ref{Fig1}, lower panel, colored lines), the angle of reflection is always identical to the incidence angle. On the other hand, the amplitude of the reflected beam generally depends on the incident polarization, the angle between $\mathbf{k}$ and $\mathbf{v}$, and the modulus of the speed, $|\mathbf{v}|$.
However, if the moving medium fulfills $\varepsilon\mu=1$, the situation is very different. First, the modified Snell's law for moving media simplifies and the angle of refraction is identical to the incident angle, as if there would be no interface. Also, this refraction angle is constant with the speed of the medium (Fig. \ref{Fig1}, lower panel, black line) and it can be proved that Lorentz invariant media are the only class of materials for which this happens (see Supplementary Material). Notwithstanding, the presence of a medium is observed in the reflected wave. When an incident circularly polarized plane-wave is reflected in a moving Lorentz invariant medium, the reflected wave presents only the opposite circular polarization component \cite{Resonant}, with an amplitude reflectivity $r= (1 - Z)/(1 + Z)$, where $Z = \sqrt{\mu/\varepsilon}$ is the relative impedance of such a medium. Therefore, for Lorentz invariant media, the reflected optical power does not depend on the speed, $\mathbf{v}$. As a result, once again, we see that Lorentz invariant media have the same properties when moving as when they are at rest.

In conclusion, we have identified a subset of materials whose electromagnetic response is the same for all inertial observers breaking the misconception that the vacuum is the only environment with such a property. Intriguingly, and to the best of our knowledge, this is the first time that such a fundamental feature of macroscopic materials satisfying $\varepsilon\mu = 1$ has been reported in the literature. In addition, we have analyzed a variety of different problems and magnitudes which confirm that Lorentz invariant materials have fundamental physical consequences over the propagation of electromagnetic waves.

We believe that the identification of Lorentz invariant materials paves the way towards a new paradigm in which the symmetries of Special Relativity will be incorporated into macroscopic electromagnetic systems, both in fundamental and applied branches of Physics. For instance, our results anticipate that interesting properties of negative refractive index metamaterials should be extensible to moving systems. Furthermore, we foresee new opportunities for analyzing electromagnetic phenomena from a symmetry-based perspective. In particular, our findings open the door to the design of electromagnetic boost eigenmodes \cite{Bliokh_LorBoostEig} and to the study of their interaction with Lorentz invariant media, based, for instance, on the symmetry breaking principle \cite{OriginKerker}. Lorentz invariant metamaterials may also have an impact on the design of photonic crystals, enlarging the set of admissible symmetries  \cite{PhotonicCrystals}.

\section*{Acknowledgements}
J.L.A. thanks Diana Castro S{\'a}nchez for her close support during the conception of the idea and the writing of the manuscript. G.M.T. acknowledges financial support from CSIC’s Quantum Technologies Platform (QTEP), from IKUR Strategy under the collaboration agreement between Ikerbasque Foundation and DIPC/MPC on behalf of the Department of Education of the Basque Government and from project PID2022-143268NB-I00 of Ministerio de Ciencia, Innovación y Universidades.

\clearpage
\newpage

\bibliography{mybib}

\begin{thebibliography}{19}%
\makeatletter
\providecommand \@ifxundefined [1]{%
 \@ifx{#1\undefined}
}%
\providecommand \@ifnum [1]{%
 \ifnum #1\expandafter \@firstoftwo
 \else \expandafter \@secondoftwo
 \fi
}%
\providecommand \@ifx [1]{%
 \ifx #1\expandafter \@firstoftwo
 \else \expandafter \@secondoftwo
 \fi
}%
\providecommand \natexlab [1]{#1}%
\providecommand \enquote  [1]{``#1''}%
\providecommand \bibnamefont  [1]{#1}%
\providecommand \bibfnamefont [1]{#1}%
\providecommand \citenamefont [1]{#1}%
\providecommand \href@noop [0]{\@secondoftwo}%
\providecommand \href [0]{\begingroup \@sanitize@url \@href}%
\providecommand \@href[1]{\@@startlink{#1}\@@href}%
\providecommand \@@href[1]{\endgroup#1\@@endlink}%
\providecommand \@sanitize@url [0]{\catcode `\\12\catcode `\$12\catcode `\&12\catcode `\#12\catcode `\^12\catcode `\_12\catcode `\%12\relax}%
\providecommand \@@startlink[1]{}%
\providecommand \@@endlink[0]{}%
\providecommand \url  [0]{\begingroup\@sanitize@url \@url }%
\providecommand \@url [1]{\endgroup\@href {#1}{\urlprefix }}%
\providecommand \urlprefix  [0]{URL }%
\providecommand \Eprint [0]{\href }%
\providecommand \doibase [0]{http://dx.doi.org/}%
\providecommand \selectlanguage [0]{\@gobble}%
\providecommand \bibinfo  [0]{\@secondoftwo}%
\providecommand \bibfield  [0]{\@secondoftwo}%
\providecommand \translation [1]{[#1]}%
\providecommand \BibitemOpen [0]{}%
\providecommand \bibitemStop [0]{}%
\providecommand \bibitemNoStop [0]{.\EOS\space}%
\providecommand \EOS [0]{\spacefactor3000\relax}%
\providecommand \BibitemShut  [1]{\csname bibitem#1\endcsname}%
\let\auto@bib@innerbib\@empty
\bibitem [{\citenamefont {Minkowski}(1908)}]{Mink_original}%
  \BibitemOpen
  \bibfield  {author} {\bibinfo {author} {\bibfnamefont {H.}~\bibnamefont {Minkowski}},\ }\bibfield  {title} {\enquote {\bibinfo {title} {Die grundgleichungen für die elektromagnetischen vorgänge in bewegten körpern},}\ }\href@noop {} {\bibfield  {journal} {\bibinfo  {journal} {Nachrichten von der Gesellschaft der Wissenschaften zu Göttingen, Mathematisch-Physikalische Klasse}\ ,\ \bibinfo {pages} {53--111}} (\bibinfo {year} {1908})}\BibitemShut {NoStop}%
\bibitem [{\citenamefont {Einstein}\ and\ \citenamefont {Laub}(1908)}]{Einstein_Laub}%
  \BibitemOpen
  \bibfield  {author} {\bibinfo {author} {\bibfnamefont {A.}~\bibnamefont {Einstein}}\ and\ \bibinfo {author} {\bibfnamefont {J.}~\bibnamefont {Laub}},\ }\bibfield  {title} {\enquote {\bibinfo {title} {On the fundamental electromagnetic equations for moving bodies},}\ }\href@noop {} {\bibfield  {journal} {\bibinfo  {journal} {Ann. Phys.}\ }\textbf {\bibinfo {volume} {26}},\ \bibinfo {pages} {532--540} (\bibinfo {year} {1908})}\BibitemShut {NoStop}%
\bibitem [{\citenamefont {Saha}\ and\ \citenamefont {Bose}(1920)}]{Minkowski}%
  \BibitemOpen
  \bibfield  {author} {\bibinfo {author} {\bibfnamefont {M.~N.}\ \bibnamefont {Saha}}\ and\ \bibinfo {author} {\bibfnamefont {S.~N.}\ \bibnamefont {Bose}},\ }\href@noop {} {\emph {\bibinfo {title} {The principle of relativity: original papers by A. Einstein and H. Minkowski.}}}\ (\bibinfo  {publisher} {Calcuta University Press},\ \bibinfo {year} {1920})\BibitemShut {NoStop}%
\bibitem [{\citenamefont {Landau}\ and\ \citenamefont {Lifshitz}(1984)}]{Landau}%
  \BibitemOpen
  \bibfield  {author} {\bibinfo {author} {\bibfnamefont {L.~D.}\ \bibnamefont {Landau}}\ and\ \bibinfo {author} {\bibfnamefont {E.~M.}\ \bibnamefont {Lifshitz}},\ }\href@noop {} {\emph {\bibinfo {title} {Electrodynamics of continuous media}}}\ (\bibinfo  {publisher} {Pergamon Press},\ \bibinfo {year} {1984})\BibitemShut {NoStop}%
\bibitem [{\citenamefont {Sommerfeld}(1952)}]{Sommerfeld}%
  \BibitemOpen
  \bibfield  {author} {\bibinfo {author} {\bibfnamefont {A.}~\bibnamefont {Sommerfeld}},\ }\href@noop {} {\emph {\bibinfo {title} {Electrodynamics. Lectures on theoretical physics vol. III}}}\ (\bibinfo  {publisher} {Academic Press Inc.},\ \bibinfo {year} {1952})\BibitemShut {NoStop}%
\bibitem [{\citenamefont {Bialynicki-Birula}\ and\ \citenamefont {Bialynicka-Birula}(2013)}]{RoleRS_Bialynicki-Birula}%
  \BibitemOpen
  \bibfield  {author} {\bibinfo {author} {\bibfnamefont {I.}~\bibnamefont {Bialynicki-Birula}}\ and\ \bibinfo {author} {\bibfnamefont {Z.}~\bibnamefont {Bialynicka-Birula}},\ }\bibfield  {title} {\enquote {\bibinfo {title} {The role of the $\text{R}$iemann–$\text{S}$ilberstein vector in classical and quantum theories of electromagnetism},}\ }\href@noop {} {\bibfield  {journal} {\bibinfo  {journal} {Journal of Physics A: Mathematical and Theoretical}\ }\textbf {\bibinfo {volume} {46}},\ \bibinfo {pages} {053001} (\bibinfo {year} {2013})}\BibitemShut {NoStop}%
\bibitem [{\citenamefont {Lakhtakia}(1994)}]{Lakhtakia_Beltrami}%
  \BibitemOpen
  \bibfield  {author} {\bibinfo {author} {\bibfnamefont {A.}~\bibnamefont {Lakhtakia}},\ }\href@noop {} {\emph {\bibinfo {title} {Beltrami fields in chiral media}}}\ (\bibinfo  {publisher} {World Scientific Co.},\ \bibinfo {year} {1994})\BibitemShut {NoStop}%
\bibitem [{\citenamefont {Pauli}(1958)}]{Pauli}%
  \BibitemOpen
  \bibfield  {author} {\bibinfo {author} {\bibfnamefont {W.}~\bibnamefont {Pauli}},\ }\href@noop {} {\emph {\bibinfo {title} {Theory of relativity}}}\ (\bibinfo  {publisher} {Pergamon Press},\ \bibinfo {year} {1958})\BibitemShut {NoStop}%
\bibitem [{\citenamefont {Veselago}(1968)}]{Veselago}%
  \BibitemOpen
  \bibfield  {author} {\bibinfo {author} {\bibfnamefont {V.~G.}\ \bibnamefont {Veselago}},\ }\bibfield  {title} {\enquote {\bibinfo {title} {The electrodynamics of substances with simultaneously negative values of $\varepsilon$ and $\mu$},}\ }\href@noop {} {\bibfield  {journal} {\bibinfo  {journal} {Sov. Phys. Usp.}\ }\textbf {\bibinfo {volume} {10}},\ \bibinfo {pages} {509--514} (\bibinfo {year} {1968})}\BibitemShut {NoStop}%
\bibitem [{\citenamefont {Pendry}(2000)}]{Pendry}%
  \BibitemOpen
  \bibfield  {author} {\bibinfo {author} {\bibfnamefont {J.~B.}\ \bibnamefont {Pendry}},\ }\bibfield  {title} {\enquote {\bibinfo {title} {Negative refraction makes a perfect lens},}\ }\href@noop {} {\bibfield  {journal} {\bibinfo  {journal} {Physical Review Letters}\ }\textbf {\bibinfo {volume} {85}},\ \bibinfo {pages} {3966--3969} (\bibinfo {year} {2000})}\BibitemShut {NoStop}%
\bibitem [{\citenamefont {Franke-Arnold}\ \emph {et~al.}(2011)\citenamefont {Franke-Arnold}, \citenamefont {Gibson}, \citenamefont {Boyd},\ and\ \citenamefont {Padgett}}]{Science_only_vacuum}%
  \BibitemOpen
  \bibfield  {author} {\bibinfo {author} {\bibfnamefont {S.}~\bibnamefont {Franke-Arnold}}, \bibinfo {author} {\bibfnamefont {G.}~\bibnamefont {Gibson}}, \bibinfo {author} {\bibfnamefont {R.~W.}\ \bibnamefont {Boyd}}, \ and\ \bibinfo {author} {\bibfnamefont {M.~J.}\ \bibnamefont {Padgett}},\ }\bibfield  {title} {\enquote {\bibinfo {title} {Rotary photon drag enhanced by a slow-light medium},}\ }\href@noop {} {\bibfield  {journal} {\bibinfo  {journal} {Science}\ }\textbf {\bibinfo {volume} {333}},\ \bibinfo {pages} {65--67} (\bibinfo {year} {2011})}\BibitemShut {NoStop}%
\bibitem [{\citenamefont {Chen}(1983)}]{Chen}%
  \BibitemOpen
  \bibfield  {author} {\bibinfo {author} {\bibfnamefont {H.~C.}\ \bibnamefont {Chen}},\ }\href@noop {} {\emph {\bibinfo {title} {Theory of electromagnetic waves: a coordinate-free approach}}}\ (\bibinfo  {publisher} {McGraw-Hill Book Co.},\ \bibinfo {year} {1983})\BibitemShut {NoStop}%
\bibitem [{\citenamefont {Fresnel}(1818)}]{Fresnel_drag}%
  \BibitemOpen
  \bibfield  {author} {\bibinfo {author} {\bibfnamefont {A.~J.}\ \bibnamefont {Fresnel}},\ }\href@noop {} {\bibfield  {journal} {\bibinfo  {journal} {Ann. Chim. Phys.}\ }\textbf {\bibinfo {volume} {9}},\ \bibinfo {pages} {57} (\bibinfo {year} {1818})}\BibitemShut {NoStop}%
\bibitem [{\citenamefont {Fizeau}(1851)}]{Fizeau_drag}%
  \BibitemOpen
  \bibfield  {author} {\bibinfo {author} {\bibfnamefont {H.}~\bibnamefont {Fizeau}},\ }\href@noop {} {\bibfield  {journal} {\bibinfo  {journal} {C. R. Acad. Sci (Paris)}\ }\textbf {\bibinfo {volume} {33}},\ \bibinfo {pages} {349} (\bibinfo {year} {1851})}\BibitemShut {NoStop}%
\bibitem [{\citenamefont {Pyati}(1967)}]{Pyati}%
  \BibitemOpen
  \bibfield  {author} {\bibinfo {author} {\bibfnamefont {V.~P.}\ \bibnamefont {Pyati}},\ }\bibfield  {title} {\enquote {\bibinfo {title} {Reflection and refraction of electromagnetic waves by a moving dielectric medium},}\ }\href@noop {} {\bibfield  {journal} {\bibinfo  {journal} {Journal of Applied Physics}\ }\textbf {\bibinfo {volume} {38}},\ \bibinfo {pages} {652} (\bibinfo {year} {1967})}\BibitemShut {NoStop}%
\bibitem [{\citenamefont {Lasa-Alonso}\ \emph {et~al.}(2023)\citenamefont {Lasa-Alonso}, \citenamefont {Olmos-Trigo}, \citenamefont {Devescovi}, \citenamefont {Hern\'andez}, \citenamefont {Garc\'{\i}a-Etxarri},\ and\ \citenamefont {Molina-Terriza}}]{Resonant}%
  \BibitemOpen
  \bibfield  {author} {\bibinfo {author} {\bibfnamefont {J.}~\bibnamefont {Lasa-Alonso}}, \bibinfo {author} {\bibfnamefont {J.}~\bibnamefont {Olmos-Trigo}}, \bibinfo {author} {\bibfnamefont {C.}~\bibnamefont {Devescovi}}, \bibinfo {author} {\bibfnamefont {P.}~\bibnamefont {Hern\'andez}}, \bibinfo {author} {\bibfnamefont {A.}~\bibnamefont {Garc\'{\i}a-Etxarri}}, \ and\ \bibinfo {author} {\bibfnamefont {G.}~\bibnamefont {Molina-Terriza}},\ }\bibfield  {title} {\enquote {\bibinfo {title} {Resonant helicity mixing of electromagnetic waves propagating through matter},}\ }\href@noop {} {\bibfield  {journal} {\bibinfo  {journal} {Phys. Rev. Res.}\ }\textbf {\bibinfo {volume} {5}},\ \bibinfo {pages} {023116} (\bibinfo {year} {2023})}\BibitemShut {NoStop}%
\bibitem [{\citenamefont {Bliokh}(2018)}]{Bliokh_LorBoostEig}%
  \BibitemOpen
  \bibfield  {author} {\bibinfo {author} {\bibfnamefont {K.~Y.}\ \bibnamefont {Bliokh}},\ }\bibfield  {title} {\enquote {\bibinfo {title} {Lorentz-boost eigenmodes},}\ }\href@noop {} {\bibfield  {journal} {\bibinfo  {journal} {Phys. Rev. A}\ }\textbf {\bibinfo {volume} {98}},\ \bibinfo {pages} {012143} (\bibinfo {year} {2018})}\BibitemShut {NoStop}%
\bibitem [{\citenamefont {Lasa-Alonso}\ \emph {et~al.}(2024)\citenamefont {Lasa-Alonso}, \citenamefont {Devescovi}, \citenamefont {Maciel-Escudero}, \citenamefont {Garc\'{\i}a-Etxarri},\ and\ \citenamefont {Molina-Terriza}}]{OriginKerker}%
  \BibitemOpen
  \bibfield  {author} {\bibinfo {author} {\bibfnamefont {J.}~\bibnamefont {Lasa-Alonso}}, \bibinfo {author} {\bibfnamefont {C.}~\bibnamefont {Devescovi}}, \bibinfo {author} {\bibfnamefont {C.}~\bibnamefont {Maciel-Escudero}}, \bibinfo {author} {\bibfnamefont {A.}~\bibnamefont {Garc\'{\i}a-Etxarri}}, \ and\ \bibinfo {author} {\bibfnamefont {G.}~\bibnamefont {Molina-Terriza}},\ }\bibfield  {title} {\enquote {\bibinfo {title} {Origin of the $\text{K}$erker phenomena},}\ }\href@noop {} {\bibfield  {journal} {\bibinfo  {journal} {Phys. Rev. Res.}\ }\textbf {\bibinfo {volume} {6}},\ \bibinfo {pages} {043311} (\bibinfo {year} {2024})}\BibitemShut {NoStop}%
\bibitem [{\citenamefont {Joannopoulos}\ \emph {et~al.}(2007)\citenamefont {Joannopoulos}, \citenamefont {Johnson}, \citenamefont {Winn},\ and\ \citenamefont {Meade}}]{PhotonicCrystals}%
  \BibitemOpen
  \bibfield  {author} {\bibinfo {author} {\bibfnamefont {J.~D.}\ \bibnamefont {Joannopoulos}}, \bibinfo {author} {\bibfnamefont {S.~G}\ \bibnamefont {Johnson}}, \bibinfo {author} {\bibfnamefont {J.~N.}\ \bibnamefont {Winn}}, \ and\ \bibinfo {author} {\bibfnamefont {R.~D}\ \bibnamefont {Meade}},\ }\href@noop {} {\emph {\bibinfo {title} {Photonic Crystals: Molding the Flow of Light}}}\ (\bibinfo  {publisher} {Princeton University Press},\ \bibinfo {year} {2007})\BibitemShut {NoStop}%
\end{thebibliography}%

\end{document}